\documentclass[12pt]{article}

\usepackage[]{graphicx,float,latexsym,times, color,cancel}
\usepackage{amsfonts,amstext,amsmath,amssymb,amsthm}
\usepackage{caption}

\long\def\symbolfootnote[#1]#2{\begingroup%
\def\thefootnote{\fnsymbol{footnote}}\footnote[#1]{#2}\endgroup}

\usepackage{titlesec} 

\titleformat{\section}{\large\bfseries}{\thesection.}{.5em}{}
\titlespacing*{\section}{0pt}{*3}{*2}
\titleformat{\subsection}{\normalfont\bfseries}{\thesubsection.}{.5em}{}
\titlespacing*{\subsection} {0pt}{*3}{*2}
\titleformat{\subsubsection}{\normalfont\bfseries}{\thesubsubsection.}{.5em}{}
\titlespacing*{\subsubsection} {0pt}{*3}{*2}

\theoremstyle{plain} 
\newtheorem{theorem}{Theorem}[section]

\theoremstyle{definition} 
\newtheorem{definition}{Definition}[section]

\renewcommand{\refname}{{\large \bf REFERENCES}}

\newcommand{\Prob} {{\bf P}}

\newcommand{\E}{{\bf E}}
\newcommand{\vv}[1]{\mathbf{#1}}

\numberwithin{equation}{section} 

\begin{document}

\title{\textbf{\Large Robust Change Detection for Large-Scale Data Streams}}

\date{}

\maketitle

\begin{center}

\author{
\null\vskip -2cm
\textbf{\large Ruizhi Zhang}\\
Department of Statistics, University of Nebraska-Lincoln, \\
Lincoln, Nebraska, USA
\vskip0.2cm
\textbf{\large Yajun Mei}\\
H. Milton Stewart School of Industrial and Systems Engineering, \\
Georgia Institute of Technology, Atlanta, Georgia, USA
\vskip0.2cm
\textbf{\large Jianjun Shi}\\
H. Milton Stewart School of Industrial and Systems Engineering, \\
Georgia Institute of Technology, Atlanta, Georgia, USA
}
\end{center}

-------------------------------------------------------------------------------------------------%

\symbolfootnote[0]{\normalsize\noindent \hspace{-6mm}Address correspondence to Ruizhi Zhang,
Department of Statistics, University of Nebraska-Lincoln,
Lincoln, Nebraska, 68583 USA;  E-mail: rzhang35@unl.edu\\
}
\\

%
{\small \noindent\textbf{Abstract:} Robust change-point detection for large-scale data streams has many real-world applications in industrial quality control, signal detection, biosurveillance. Unfortunately, it is highly non-trivial to develop efficient schemes due to three challenges: (1) the unknown sparse subset of affected data streams, (2)  the unexpected outliers, and (3) computational scalability for real-time monitoring and detection. In this article, we develop a family of efficient real-time robust detection schemes for monitoring large-scale independent data streams. For each data stream, we propose to construct a new local robust detection statistic called $L_{\alpha}$-CUSUM statistic that can reduce the effect of outliers by using the Box-Cox transformation of the likelihood function. Then the global scheme will raise an alarm based upon the sum of the shrinkage transformation of these local $L_{\alpha}$-CUSUM statistics so as to filter out unaffected data streams. In addition,  we propose a new concept called {\em false alarm breakdown point} to measure the robustness of online monitoring schemes and propose a {\em worst-case detection efficiency score} to measure the detection efficiency when the data contain outliers. We then characterize the breakdown point and the efficiency score of our proposed schemes.   Asymptotic analysis and numerical simulations are conducted to illustrate the robustness and efficiency of our proposed schemes.}

\section{Introduction}

Robust statistics have been extensively studied in the offline context when the entire data set  is available for decision-making and is contaminated with outliers, e.g., robust estimation \cite{huber:1964,basu:1998}, robust hypothesis testing \cite{huber:1965, heritier:1994}, and robust regression \cite{yohai:1987,cantoni:2001}. Also, see the classical books, \cite{huber:2009} or \cite{hampel:2011},  for literature review.
In this paper, we propose to develop robust methods in the context of  sequential change-point detection when one is interested in detecting sparse, persistent smaller changes in large-scale data streams under the contamination of transient larger outliers. The problem of robust monitoring large-scale data streams in the presence of outliers  occurs in many real-world applications such as industrial quality control, biosurveillance, key infrastructure, or internet traffic  monitoring, in which sensors are deployed to constantly monitor the changing environment, see  \cite{shmueli:2010},\cite{tartakovsky:2013},\cite{yan:2015}. Unfortunately, it  is highly non-trivial to develop efficient, robust real-time monitoring schemes or algorithms due to three challenges: (1) the sparsity, where only a few unknown data streams might be affected; (2) the robustness, where we are interested in detecting smaller persistent changes, not the larger transient outliers; and
(3) the computational scalability, where the algorithms can be implemented recursively to make real-time decisions.

In the literature of sequential change-point detection for  a large number of data streams, to the best of our knowledge, while the sparsity issue has been investigated, no research has been done on the robustness issue. To be more specific, the sparsity has been first addressed by \cite{xie:2013} using a semi-Bayesian approach and later by \cite{Wang:2015} using shrinkage-estimation-based schemes. \cite{chan:2015} developed asymptotic optimality theory for large-scale independent Gaussian data streams.  Unfortunately, all these methods are sensitive to outliers since they are based on the likelihood function of specific parametric models (e.g. Gaussian) of the observations.  Meanwhile, regarding the robustness issue, research is available for monitoring one- or low- dimensional streaming data such as rank-based method in \cite{gordon:1994,gordon:1995}, kernel-based method in \cite{desobry:2005}. However, these nonparametric methodologies generally lose detection efficiency under specific parametric or semi-parametric models. By considering the worst-case of the outlier distribution,  \cite{unnikrishnan:2011} formulated the problem of finding the optimal robust change detection procedure by solving a minimax problem. However, the resulting optimal test is based on the  least-favorable-pair distributions of two uncertainty sets, which depends on the information of outliers.  More importantly, it is unclear how to extend their method from monitoring a single data stream to  monitoring multiple data streams when we also need to deal with the sparsity issue in which there is uncertainty on the subset of affected data streams.



In this paper, we develop efficient real-time monitoring schemes that are able to robustly detect smaller persistent changes in the presence of  larger transient outliers when online monitoring of large-scale data streams. From the methodology viewpoint, our proposed schemes are semi-parametric and extend two contemporary concepts to the context of online monitoring of data streams: (i)  $L_{q}$-likelihood \cite{ferrari:2010, qin:2016} for robustness, and (ii) the sum-shrinkage technique  \cite{liu:2017,zhang2018asymptotic} for sparsity. These allow us to develop statistically efficient and computationally simple schemes  that can be implemented recursively over time for robust real-time monitoring of a large number of data streams. Moreover, we also extend the concept of breakdown in the offline robust statistics \cite{hampel:1968} to the sequential change-point detection context and conduct  the false alarm breakdown point analysis, which turns out to be useful for  the choices of tuning parameters in our proposed schemes.

We should point out that our  contribution is not on the optimality theory  but on the asymptotic properties of our proposed schemes that include the classical CUSUM-based procedures as a special case. Our research makes four contributions in the statistics field by combining robust statistics with sequential change-point detection for large-scale data streams.  First, our proposed method is robust to infrequent outliers as well as the uncertainty of affected data streams.
Second, our proposed method can be implemented recursively and distributed via parallel computing and thus is suitable for real-time monitoring over a long time period.
Third, inspired by the concept of breakdown point \cite{hampel:1968} in the offline robust statistics, 
we propose a novel concept of false alarm breakdown point to  quantify the robustness of any online monitoring schemes and show that our proposed schemes  indeed have much larger false alarm breakdown point than the classical CUSUM-based schemes. 
Finally, from the mathematical viewpoint, we use Chebyshev's inequality to derive non-asymptotic lower bounds on the average run length of false alarm of our proposed methods. The non-asymptotic results hold regardless of dimensionality and allow us to provide a deep insight into the effect of high-dimensionality in change-point detection under the modern asymptotic regime when the dimension or the number of data streams goes to $\infty.$


The remainder of this article is organized as follows. In Section \ref{sec:form}, we start with problem formulations and model assumptions.  In Section \ref{sec:method}, we introduce our proposed family of robust monitoring schemes. In Section \ref{sec:dae}, the properties of the detection efficiency of our proposed schemes and  the guideline to choose tuning parameters in our proposed schemes are provided. Then, we investigate the robustness of our proposed methods by conducting breakdown point analysis in Section \ref{sec:bd}. Simulation results are presented in Section \ref{sec:simu}.  In Section \ref{sec:con}, we conclude our paper with a few remarks The proofs of our main theorems are postponed to Appendix.

\section{Problem Formulation}\label{sec:form}

Suppose we are monitoring  $K$  independent data streams in a system.
\begin{eqnarray} \label{eqn01}
\mbox{Data Stream $1:$} && X_{1,1}, X_{1,2}, \cdots \\
\mbox{Data Stream $2:$} && X_{2,1}, X_{2,2}, \cdots \cr
\ldots && \ldots  \cr
\mbox{Data Stream $K:$} && X_{K,1}, X_{K,2}, \cdots. \nonumber
\end{eqnarray}

Under the classical change-point detection model for monitoring multi-streams (e.g., \cite{xie:2013,chan:2015,fellouris2016second,liu:2017,zhang2018asymptotic}), one assumes that the data $X_{k,n}$'s are initially independent and identically distributed (i.i.d.) with probability density function (pdf) $f_0(x)$. At some unknown time $\nu \ge 1,$ an undesired event occurs, and change the distributions of  $m$ out of $K$ data streams, i.e., the affected local streams $X_{k,n}$'s have another distribution $f_1(x)$ when $n \ge \nu.$ The objective is to raise an alarm as soon as possible once a change occurs. Here, we refer to this classical model as the idealized model.

In this paper, we investigate the change-point detection problem under Tukey-Huber's gross error model. As mentioned in the introduction, we want to raise an alarm as quickly as possible if there is a persistent distribution change on the data, but we prefer to take observations without any actions if there are only transient outliers. Mathematically, we assume the distribution of data $X_{k,n}$ might be changed from $h_{0}$ to $h_{1}$ at some change time $\nu,$  the $h_0$ and $h_1$ are the Tukey-Huber's gross error model of the mixture densities
\begin{eqnarray} \label{equ:highgrosserror}
h_0(x) = (1-\epsilon) f_{0}(x)+\epsilon g_0(x), \quad
h_1(x) = (1-\epsilon) f_{1}(x)+\epsilon g_1(x),
\end{eqnarray}
where $\epsilon\in [0,1)$  is referred to as the contamination/outlier ratio, $g_0$ and $g_1$ are the (unknown) outlier distributions.
Denote by $\Prob_{h_0}^{(\infty)}$ and $\E_{h_0}^{(\infty)}$ the probability measure and expectation when the data $X_{k,n}$'s are i.i.d. with the density $h_0$ when no change occurs, and denote by $\Prob_{h_1}^{(\nu)}$ and $\E_{h_1}^{(\nu)}$ the same when the change occurs at time $\nu$ and $m$ out of $K$ streams $X_{k,n}$'s have the post-change distribution $h_1$. 

As in the classical sequential change-point problem, a statistical procedure under our setting is defined as a stopping time $T$ that represents the time when we raise an alarm to declare that a change has occurred. Here $T$ is an integer-valued random variable, and the decision $\{T=t\}$ is based only on the observations in  the first $t$ time steps. To evaluate the performance of the detection procedure $T$ under Tukey-Huber's gross error model when the outlier distributions $g_0$ and $g_1$ in (\ref{equ:highgrosserror}) are unknown, we first assume the average run length to false alarm of the procedure $T$ is controlled under the idealized model. That is, we assume that the procedure $T$ is designed to satisfy the false alarm constraint


\begin{eqnarray} \label{equ:constr2}
\E^{(\infty)}_{f_{0}}(T) \ge \gamma,
\end{eqnarray}
for some pre-specified value $\gamma > 0.$ We then investigate the robustness and the detection efficiency of the monitoring procedure under the gross error model in (\ref{equ:highgrosserror}).

First, we propose quantifying  the robustness of a monitoring procedure $T$ under the gross error model in (\ref{equ:highgrosserror}) by borrowing the concept of breakdown point analysis from the offline robust statistics literature. To be more specific, we propose to define a new concept called false alarm breakdown point, which characterizes the minimal percentage of outliers that can make the false alarm rate under the gross error model $h_{0}= (1-\epsilon) f_{0}(x)+\epsilon g_0(x)$ very different from that under the idealized model $f_{0}.$ 

{\em The false alarm breakdown point} $\epsilon^*(T)$ of a family of monitoring schemes $T(b)$'s is defined as
\begin{eqnarray}\label{breakdownl}
\epsilon^{*}(T)= \inf \{\epsilon \ge 0:\ \underset{h_{0} \in \hbar_{0,\epsilon}}{\inf} \log (\E^{(\infty)}_{h_{0}}T(b_{\gamma}))=o(\log\gamma)\},
\end{eqnarray}
where $\E^{(\infty)}_{f_{0}}(T(b_{\gamma})) \sim \gamma$  as $\gamma \to \infty$,
and the  set $\hbar_{0,\epsilon}$ is the $\epsilon$-contaminated distribution density class of the idealized model $f_{0}(x)$ for given $\epsilon\in[0,1):$
\begin{eqnarray}\label{e-neigh}
\hbar_{0,\epsilon}=\{h|h=(1-\epsilon)f_{0}+\epsilon g, g\in \mathcal G\},
\end{eqnarray}
and   $\mathcal G$ denotes the class of all probability densities of the data $X_{k,n}$.

Roughly speaking, the false alarm breakdown point characterizes the minimal percentage of outliers that can make the designed average run length to false alarm  $\gamma$ unreliable. Thus, a scheme with larger breakdown points is more robust.


Second, we quantify the detection efficiency of the monitoring procedure $T$ under the gross error model in (\ref{equ:highgrosserror}). For that purpose, recall that under the Lorden's minimax criteria  \cite{lorden:1971},  the  worst-case detection delay under $h_1$  is defined as 
 \begin{eqnarray} \label{equ:delay3}
\vv D_{h_1}(T) = \quad  \sup_{\nu \ge 1}\
\mbox{ess}\sup \E_{h_1}^{(\nu)} \left( (T - \nu + 1)^{+} \big| \mathcal{F}_{\nu-1} \right).
\end{eqnarray}
Here  $\mathcal{F}_{\nu-1} = (X_{1,[1,\nu-1]}, \ldots, X_{K, [1,
\nu-1]})$ denotes past global information at time $\nu.$
$X_{k,[1,\nu-1]}= (X_{k,1}, \ldots, X_{k,\nu-1})$ is past local
information for the $k$-th data stream.
However, since the outlier distribution $g_1$ is unknown, 
we propose to define two quantities on detection efficiency: one is {\em asymptotic efficiency score} defined by
\begin{eqnarray}\label{eq:ae}
\text{AE}(T, \epsilon; g_0, g_1)=\lim_{\gamma\to\infty}\frac{\log(\E^{(\infty)}_{h_0}(T(b_{\gamma})))}{\vv D_{h_1}(T(b_{\gamma})) },
\end{eqnarray}
and the other is the {\em worst-case asymptotic efficiency score} defined by
\begin{eqnarray}\label{eq:wae}
\text{WAE}(T, \epsilon)=\underset{g_0\in \mathcal G, g_1\in \mathcal G}{\inf}\text{AE}(T, \epsilon; g_0, g_1)=\lim_{\gamma\to\infty}\frac{\underset{g_0\in \mathcal G}{\inf}\left[\log(\E^{(\infty)}_{h_0}(T(b_{\gamma})))\right]}{\underset{g_1\in \mathcal G}{\sup}\left[\vv D_{h_1}(T(b_{\gamma}))\right] }.
\end{eqnarray}

In both definitions, $b_{\gamma}$ is a threshold of $T = T(b_{\gamma})$ so that $\E^{(\infty)}_{f_0}(T(b_{\gamma}))\sim \gamma.$ Clearly, when the data contain outliers, the procedure with a larger asymptotic efficiency score implies more efficiency in detecting the persistent change.
Note that the definition of the  asymptotic efficiency  in (\ref{eq:ae}) depends on the outlier distributions $g_0$ and $g_1,$  which are unknown in practice,
but the worst-case detection efficiency  $\text{WAE}(T,\epsilon)$ in (\ref{eq:wae}) measures the worst case among all set $\mathcal G$ of outlier distributions $g_0$ and $g_1.$ Note when  we are monitoring a single data stream, i.e., the dimension $K=1,$ the optimal procedure that maximizes $\text{WAE}(T, \epsilon)$ is a CUSUM procedure constructed by a least-favorable-pair $g^*_0, g^*_1,$ as shown in \cite{unnikrishnan:2011}. However, the problem of finding the optimal procedure that minimizes $\text{WAE}(T, \epsilon)$ becomes more complicated when the dimension $K$ is large and the set of affected data streams is unknown.


In this paper, our objective is to develop a family of efficient, robust monitoring schemes that have a large breakdown point $\epsilon^*(T)$ in (\ref{breakdownl}) and  a large worst-case asymptotic efficiency score $\text{WAE}(T,\epsilon)$ in (\ref{eq:wae})  subject to the constraints that this family of schemes satisfy the false alarm constraint in (\ref{equ:constr2}) under the idealized pre-change distribution $f_0.$

\section{Our proposed method}\label{sec:method}

In this section, we will present our proposed schemes.  At the high-level, our proposed schemes include two components: (i) robust monitoring each local data stream individually in parallel, and then (ii) combining local detection statistics to make an online global-level decision. For the purpose of easy understanding, we split the presentation of our proposed schemes into two subsections, and each subsection focuses on one component  of the proposed scheme.

\subsection{Robust local statistics}


For the $k^{th}$ data stream, we propose to define a new local $L_{\alpha}$-CUSUM statistic:
\begin{eqnarray} \label{eqLalpha}
W_{\alpha,k, n}=\max\Big( W_{\alpha,k, n-1} +  \frac{[f_{1}(X_{k,n})]^{\alpha}-[f_{0}(X_{k,n})]^{\alpha}}{\alpha}, 0\Big),
\end{eqnarray}
for $n \ge 1,$ and $W_{\alpha,k, 0}= 0.$ Here $\alpha \ge 0$ is a tuning parameter that can control the tradeoff between statistical efficiency and robustness under the gross error model in (\ref{equ:highgrosserror}) and its suitable choice will be discussed later.

The motivation of our $L_{\alpha}$-CUSUM statistic in (\ref{eqLalpha}) is as follows. Recall that when locally monitoring the single $k^{th}$ data stream $X_{k,n}$ with a possible local distribution change from $f_{0}$ to $f_{1},$ the generalized likelihood ratio test becomes the classical CUSUM statistic $W_{k, n}^{*},$ which has a recursive form:
\begin{eqnarray} \label{eq3n011}
W^*_{k, n} = \max_{1 \le \nu < \infty} \log\frac{\prod_{i=1}^{\nu-1} f_{0}(X_{k,i}) \prod_{i=\nu}^{n} f_{1}(X_{k,i})}{\prod_{i=1}^{n} f_{0}(X_{k,i})}
= \max\Big( W^*_{k, n-1} +
\log \frac{f_{1}(X_{k,n})}{f_{0}(X_{k,n})},\ 0\Big).
\end{eqnarray}
The CUSUM statistic enjoys nice optimality properties when all models are fully correctly specified \cite{moustakides:1986}, but unfortunately  it is very sensitive to the outliers as in all other likelihood based methods in offline statistics. One recent idea in offline robust statistics is to replace the log-likelihood statistic $\log f(X)$  by $L_{\alpha}$-likelihood statistic $([f(X)]^{\alpha}-1)/\alpha$ for some $\alpha > 0,$ see \cite{ferrari:2010},\cite{qin:2016}. At the high-level, $L_{\alpha}$-likelihood statistic $\frac{[f(X)]^{\alpha} - 1}{\alpha}$ is always bounded below by $-1/\alpha$ whereas  the log-likelihood statistic $\log f(X)$ could go to $-\infty$. Thus, the impact of outliers is bounded for the $L_{\alpha}$-likelihood statistic but unbounded for the log-likelihood statistic. Moreover, as $\alpha \to 0,$ the $L_{\alpha}$-likelihood function converges to the log-likelihood statistic, and thus it keeps statistical efficiencies when $\alpha$ is small. Here we apply this idea to develop our $L_{\alpha}$-CUSUM statistic. More rigorous robust properties  will be discussed later in Section \ref{sec:bd}.

%

\subsection{Efficient global monitoring statistics}

With local $L_{\alpha}$-CUSUM statistics $W_{\alpha,k,n}$ in (\ref{eqLalpha}) for each local stream, it is important to fuse these local statistics together smartly so as to address the sparsity issue. Here we propose to combine these local statistics together via the sum-shrinkage technique in \cite{liu:2017}, i.e., we raise a global-level alarm at time

\begin{eqnarray}\label{shrinkage}
N_{\alpha}(b)=\inf\left\{n \ge 1: \sum_{k=1}^{K}h(W_{\alpha,k,n})\ge b\right\},
\end{eqnarray}
where $h(\cdot) \ge 0$ are some suitable shrinkage transformation functions, and $b > 0$ is a pre-specified constant. Intuitively,  the shrinkage functions $h(\cdot)$'s in (\ref{shrinkage}) play the role of dimension reduction by automatically filtering out those non-changing local data streams and by keeping only those local streams that might provide information about the changing event. This will allow us to improve the detection power in the sparsity scenario when only a few local features are involved in the change.


For the purpose of illustration, here we focus on two kinds of  shrinkage functions: one is the soft-thresholding function $h(x) = \max\{x-d, 0\},$ and the other is the order-thresholding function $h(x) = x {\bf 1}\{x \ge w_{(r)}\},$ where $w_{(r)}$ is the $r$-th largest statistic of $w_1, \cdots, w_{K}.$
Then the corresponding two global monitoring schemes are defined by
\begin{eqnarray}
N^{(soft)}_{\alpha}(b,d)&=&\inf\left\{n \ge 1: \sum_{k=1}^{K}\max\{0, W_{\alpha,k,n}-d\}\ge b\right\}, \label{soft}\\
N^{(r)}_{\alpha}(b) &=& \inf\Big\{n \ge 1: \sum_{k=1}^{r} W_{\alpha,(k),n} \ge b \Big\}, \label{rankCUSUM}
\end{eqnarray}
where $W_{\alpha, (1), n} \ge W_{\alpha, (2), n} \ge \ldots \ge W_{\alpha, (K),n}$ are the order statistics of the $K$ local $L_{\alpha}$-CUSUM statistics $W_{\alpha,1,n}, \ldots, W_{\alpha, K,n}.$

%

One can also consider  other shrinkage functions such as the detectability score tansformation $h(x)=\log \left[1-p_0+0.64 p_0 \exp(x/2)\right]$ proposed in \cite{chan:2015}. This yields another global monitoring scheme
\begin{eqnarray} \label{eqn_Chana}
 N_{Chan,\alpha}(b, p_0)=\inf\left\{n \ge 1:  \sum_{k=1}^{K} \log \left[1-p_0+0.64*p_0 \exp(W_{\alpha,k,n}/2)\right]  \ge b \right\}.\nonumber
\end{eqnarray}
Our extensive numerical simulation experiences illustrate that for a given $\alpha,$ the scheme $N_{Chan,\alpha}(b)$ in (\ref{eqn_Chana}) has the similar statistical/robustness properties to those schemes $N^{(soft)}_{\alpha}(b,d)$ and $N^{(r)}_{\alpha}(b)$ in (\ref{soft})  and (\ref{rankCUSUM}) in many interesting sparse post-change scenarios when $p_0 = r/K.$ This is because all these procedures utilize the same local $L_{\alpha}$-CUSUM statistics $W_{\alpha,k,n}$ in (\ref{eqLalpha}) and aim to detect the same post-change scenarios (after regularization).

Besides these aforementioned shrinkage transformations, there are other approaches to combine the local detection statistics together to make a global alarm. Two popular  approaches in the literature are the ``MAX" and the ``SUM" schemes, see \cite{tartakovsky:2008} and \cite{mei:2010}:
\begin{eqnarray} \label{eqnmax}
N_{\alpha,\max}(b) &=& \inf\left\{n \ge 1: \max_{1 \le k \le K} W_{\alpha,k,n} \ge b \right\}, \\
N_{\alpha,\text{sum}}(b) &=& \inf\left\{n \ge 1: \sum_{k=1}^{K} W_{\alpha,k,n} \ge b \right\}.  \label{eqnsum}
\end{eqnarray}
On one hand, the ``MAX" and the ``SUM" schemes could be considered as the special cases of our proposed top-r based scheme $N^{(r)}_{\alpha}(b)$ in (\ref{rankCUSUM}) when $r=1$ and $r=K$ respectively. On the other hand, the ``MAX" and ``SUM" approaches are generally statistically inefficient unless in extreme cases of  very few or many affected local data streams.

Note that there are three tuning parameters in our proposed schemes: $(\alpha, d, b)$ for the schemes $N^{(soft)}_{\alpha}(b,d)$ in (\ref{soft}) and  $(\alpha, r, b)$ for the scheme $N^{(r)}_{\alpha}(b)$ in (\ref{rankCUSUM}). It is natural to ask what  are the ``optimal" choices of these tuning parameters. It turns out that the most challenging one is the optimal choice of the common parameter $\alpha$, which is related
to the robustness  from the gross error models in (\ref{equ:highgrosserror}), and will be discussed in Section \ref{sec:bd}. Next, the ``optimal" choice of the shrinkage parameter $d$ or $r$ mainly depends on the number of affected local data streams, see our asymptotic properties in the next section. Finally, the choice of the threshold $b$ is straightforward for given two other parameters since it can be chosen to satisfy the false alarm constraint in (\ref{equ:constr2}).

\section{Worst-case asymptotic efficiency score}\label{sec:dae}

In this section, we derive the worst-case asymptotic efficiency score (\ref{eq:wae}) of our proposed schemes $N^{(soft)}_{\alpha}(b,d)$ in (\ref{soft}) and $N^{(r)}_{\alpha}(b)$ in (\ref{rankCUSUM}). To see that, we first
report two standard change-point detection properties of our proposed schemes: the ARL to false alarm and detection delay under the gross error model $h_{i} = (1-\epsilon) f_{i} + \epsilon g_i,$ where the outlier distributions $g_i$ are given and $i=0,1.$ Then, we will look at the worst-case of the outlier distributions to derive the worst-case asymptotic efficiency. It is important to note that our proposed schemes do not involve the contamination ratio $\epsilon$ or the information of outliers $\epsilon, g.$ Finally, based on our detection delay analysis, we provide guidelines on how to choose the tuning parameters in our proposed schemes.
The proofs of the theorems are presented in the Appendix.


Let us begin with the definition of the expectation of the $L_\alpha$-likelihood ratio statistic $Y = ([f_{1}(X)]^{\alpha}-[f_{0}(X)]^{\alpha})/ \alpha$ when $X$ is distributed according to $h_{i}= (1-\epsilon) f_{i} + \epsilon g_i$ for given $\epsilon,$ $g_i,$ and $i=0,1$ Note that when $\alpha = 0,$ the variable $Y $ should be treated as the log-likelihood ratio $\log(f_{1}(X)/f_{0}(X))$.
\begin{definition}\label{assumptionsection3no2}
Given $\epsilon\ge 0$ and $\alpha \geq 0,$  for $i=0,1,$ define
\begin{eqnarray} \label{conditionful00}
I_{i}(\epsilon,\alpha;g_i)&=&\E_{h_{i}}\Big[\frac{[f_{1}(X)]^{\alpha}-[f_{0}(X)]^{\alpha}}{\alpha}\Big]\\
&=&(1-\epsilon)\E_{f_{i}}\Big[\frac{[f_{1}(X)]^{\alpha}-[f_{0}(X)]^{\alpha}}{\alpha}\Big]
+ \epsilon\E_{g_i}\Big[\frac{[f_{1}(X)]^{\alpha}-[f_{0}(X)]^{\alpha}}{\alpha}\Big].\nonumber
\end{eqnarray}
\end{definition}
Note when $\epsilon=0,$ $I_{i}(\epsilon=0,\alpha;g_i)$ does not depend on $g_i.$ So we further denote $I_{i}(\alpha):=I_{i}(\epsilon=0,\alpha;g_i)$ for simplification.
It turns out that the ARL to false alarm and detection delay of our proposed schemes are depending on whether $I_{i}(\epsilon,\alpha;g_i) < 0$ or $> 0.$
Next, let us summarize the false alarm properties of our proposed schemes under the gross error model $h_{0}= (1-\epsilon) f_{0} + \epsilon g_0.$
\begin{theorem}\label{thm1}
Assume $I_{0}(\epsilon,\alpha;g_0)<0,$ then there exists a unique positive constant $\lambda(\epsilon, \alpha; g_0)$ depends on $f_{0}, f_{1}, g_0,\alpha, \epsilon$ such that
\begin{eqnarray} \label{constantlambda}
 \E_{h_{0}} \exp\Big\{\lambda(\epsilon,\alpha; g_0) \frac{[f_{1}(X)]^{\alpha}-[f_{0}(X)]^{\alpha}}{\alpha}\Big\}= 1.
\end{eqnarray}
With the constant $\lambda(\epsilon,\alpha; g_0)>0$ in (\ref{constantlambda}),  the ARL to false alarm of our proposed schemes, $N^{(soft)}_{\alpha}(b,d)$ in (\ref{soft}) and $N^{(r)}_{\alpha}(b)$ in (\ref{rankCUSUM}),  are given as follows under different sufficient conditions:

\noindent
(a) When $\lambda(\epsilon,\alpha; g_0) b > K \exp\{- \lambda(\epsilon,\alpha; g_0)d\},$ we have
  \begin{eqnarray}\label{a}
\E_{h_0}^{(\infty)}[N^{(soft)}_{\alpha}(b,d)] \ge  \frac14  \exp\left( \left[\sqrt{\lambda(\epsilon,\alpha; g_0) b}-\sqrt{K \exp\{- \lambda(\epsilon,\alpha; g_0)d\}}\right]^2\right).
\end{eqnarray}
(b) When $\lambda(\epsilon,\alpha; g_0) b > K,$ we have
  \begin{eqnarray}\label{aa}
\E_{h_0}^{(\infty)}[N^{(r)}_{\alpha}(b)] \ge   \frac14  \exp\left( \left[\sqrt{\lambda(\epsilon,\alpha; g_0) b}-\sqrt{K}\right]^2\right).
\end{eqnarray}

\end{theorem}


Let us add some comments to better understand the theorem. First, the existence of the unique constant $\lambda(\epsilon,\alpha; g_0)>0$ in (\ref{constantlambda}) is based on the assumption that $I_{0}(\epsilon,\alpha;g_0)<0$, see Appendix A2 of \cite{wald1973sequential}. Moreover, when $\epsilon=0,$ both $I_{0}(0,\alpha;g_0)$ and $\lambda(0,\alpha; g_0)$ only depend on the idealized model $f_{i}(x)$ and $\alpha,$ but do not depend on the information of outliers, i.e., $\epsilon$ and $g_i.$ For simplification, we denote $\lambda(\alpha)=\lambda(0,\alpha; g_0).$

Second, our rigorous, non-asymptotic results in (\ref{a}) and (\ref{aa}) hold no matter how large the number $K$ of data streams is. This allows us to investigate the modern asymptotic regime when the dimension $K$ goes to $\infty.$

Finally, the assumptions of $\lambda(\epsilon,\alpha; g_0) b > K \exp\{- \lambda(\epsilon,\alpha; g_0)d\}$ or $\lambda(\epsilon,\alpha; g_0) b > K$ essentially says that the global threshold $b$ of our proposed schemes should be large enough if one wants to control the global false alarm rate when online monitoring large-scale streams. These results allow us to find a conservative threshold $b$ so as to satisfy the false alarm constraint in (\ref{equ:constr2}), also see the details of parameter setting below.

%

Next, the following theorem summarizes  the detection delays of our proposed schemes $N^{(soft)}_{\alpha}(b,d)$ in (\ref{soft}) and $N^{(r)}_{\alpha}(b)$ in (\ref{rankCUSUM})  when $m$ out of $K$ features are affected by the occurring event for some given $1 \le m \le K.$ The detailed proof of Theorem \ref{thm1b} will be presented in Appendix.

\begin{theorem}\label{thm1b}
 Suppose $I_{1}( \epsilon,\alpha;g_1) > 0$ , and  $m$ out of $K$ features are affected.

 \noindent
(a) If $b/m + d$ goes to $\infty$, then the detection delay of $N^{(soft)}_{\alpha}(b,d)$ satisfies
\begin{eqnarray}\label{bound}
\vv D_{h_1}(N^{(soft)}_{\alpha}(b,d))\le (1+o(1)) \frac{1}{I_{1}( \epsilon,\alpha;g_1)} \left( \frac{b}{m}+d\right),
\end{eqnarray}
(b) If $r\ge m$ and $b/m$ goes to $\infty$, then the detection delay of $N^{(r)}_{\alpha}(b)$ satisfies
\begin{eqnarray}\label{bound}
\vv D_{h_1}(N^{(r)}_{\alpha}(b))\le (1+o(1)) \frac{1}{I_{1}( \epsilon,\alpha;g_1)} \left( \frac{b}{m}\right),
\end{eqnarray}
where the $o(1)$ term does not depend on the dimension $K$, but might depend on $m$ and $\alpha$ as well as the distributions $h_{1}.$
\end{theorem}

To simplify the notation, we use $N_{\alpha}$ to denote both the scheme $N^{(soft)}_{\alpha}(b,d)$ and scheme $N^{(r)}_{\alpha}(b).$ By Theorem \ref{thm1} and Theorem \ref{thm1b}, when $K$ is fixed, if $I_{0}( \epsilon,\alpha;g_0)<0$ and $I_{1}( \epsilon,\alpha;g_1)>0,$ we can get a natural lower bound of the asymptotic efficiency score (\ref{eq:ae}) of our proposed schemes,
\begin{eqnarray}
\text{AE}(N_{\alpha},\epsilon;g_0,g_1)\ge m\lambda(\epsilon,\alpha;g_0)I_{1}( \epsilon,\alpha;g_1).
\end{eqnarray}
However, if we can find outlier distributions $g^*_0, g^*_1$ such that $I_{0}( \epsilon,\alpha;g^*_0)>0$ and $I_{1}( \epsilon,\alpha;g^*_1)<0,$ we will get
\begin{eqnarray}
\text{AE}(N_{\alpha},\epsilon;g^*_0,g^*_1)=0,
\end{eqnarray}
which implies the procedure cannot detect the persistent change from $f_0$ to $f_1$ at all due to the contamination of outliers.

Now, we are ready to present the worst-case asymptotic efficiency score of our proposed scheme $N_{\alpha}.$ First, assume $I_0(\alpha)=\E_{f_{0}}\Big[\frac{[f_{1}(X)]^{\alpha}-[f_{0}(X)]^{\alpha}}{\alpha}\Big]<0$ and $I_1(\alpha)=\E_{f_{1}}\Big[\frac{[f_{1}(X)]^{\alpha}-[f_{0}(X)]^{\alpha}}{\alpha}\Big]>0,$  denote
\begin{eqnarray}
M^*(\alpha)=\underset{x}{{\rm ess} \sup } |\frac{[f_{1}(x)]^{\alpha}-[f_{0}(x)]^{\alpha}}{\alpha}|.
\end{eqnarray}
Then we have the following theorem:
\begin{theorem}\label{the:wae}
For our proposed scheme $N_{\alpha}(b)$with $\alpha\ge 0,$ suppose $K$ is fixed and  $b\to \infty,$

\noindent
(a) if $\epsilon<-I_0(\alpha)/[M^*(\alpha)-I_0(\alpha)]$ and $\epsilon<I_1(\alpha)/[M^*(\alpha)+I_1(\alpha)],$ we have
\begin{eqnarray}
\text{WAE}(N_{\alpha},\epsilon)&\ge&m\lambda^*(\epsilon,\alpha)\Big[(1-\epsilon)I_1(\alpha)-\epsilon M^*(\alpha)\Big]>0,
\end{eqnarray}
where $\lambda^*(\epsilon,\alpha)=\underset{g_0\in \mathcal G}{\inf}\lambda(\epsilon,\alpha;g_0)>0.$

\noindent(b) Otherwise, $\text{WAE}(N_{\alpha},\epsilon)=0.$
\end{theorem}

Note if $\log (f_1(x)/f_0(x))$ is unbounded, we have $M^*(0)=+\infty.$ Based on Theorem \ref{the:wae},  for any $\epsilon>0, \text{WAE}(N_{\alpha=0},\epsilon)=0,$ which implies the CUSUM based method cannot detect the persistent change at all under any percentage of outliers. However, if both $f_0, f_1$ are bounded, for any $\alpha>0,$ we have $M^*(\alpha)<+\infty.$  Thus, our proposed schemes $N_{\alpha}$ will always have a positive worst-case asymptotic efficiency score when the contamination ratio $\epsilon$ is small. This implies the detection efficiency of our proposed schemes under the gross error model.

Note that there are three tuning parameters in our proposed schemes: $(\alpha, d, b)$ for the schemes $N^{(soft)}_{\alpha}(b,d)$ in (\ref{soft}) and  $(\alpha, r, b)$ for the scheme $N^{(r)}_{\alpha}(b)$ in (\ref{rankCUSUM}). It is natural to ask what  are the ``optimal" choices of these tuning parameters. It turns out that Theorems \ref{thm1} and \ref{thm1b} provide the optimal choices of $(d,b)$ or $(r,b)$ that asymptotically minimize the detection delay subject to the false alarm constraint $\gamma$ in (\ref{equ:constr2}). Below we will report the corresponding results, and detailed proofs and descriptions are postponed to Appendix.

(1) The optimal choice of parameter $\alpha$, which turns out to be the most challenging one, as it is related to the robustness  from the gross error models in (\ref{equ:highgrosserror}). We will discuss in more details in Section \ref{sec:bd} through the concept of false alarm breakdown point. The result in Section  \ref{sec:bd} shows the optimal $\alpha_{opt}$ only depends on the distributions $f_{0},f_{1}$ but independent of other parameters $d,r,b,$ and outliers information $\epsilon,g.$

(2) Given $\alpha_{opt},$ the choice of the shrinkage parameter $d$ or $r$ mainly depends on the number $m$ of affected local feature coefficients. If we want to minimize the detection delay subject to the false alarm constraint $\gamma$ in (\ref{equ:constr2}),  we can set $r =m $ for the scheme $N^{(r)}_{\alpha_{opt}}(b)$ in (\ref{rankCUSUM}). The optimal choice of $d$ for the proposed scheme
$N^{(soft)}_{\alpha_{opt}}(b,d)$ in (\ref{soft}) is a little complicated, and given by
 \begin{eqnarray}\label{d1}
d_{opt}=\frac{1}{\lambda(\alpha_{opt})}\left(\log\frac{K}{m} +\log\frac{\log\gamma}{m}   \right),
\end{eqnarray}
where $\lambda( \alpha_{opt})$ is defined in (\ref{constantlambda}) and only depends on $f_{0}, f_{1}$ and $\alpha_{opt}.$

(3) The choice of the threshold $b$ is straightforward for given two other parameters, since it can be chosen to satisfy the false alarm constraint in (\ref{equ:constr2}) under the idealized distribution $f_{0}.$ A choice of global detection threshold
\begin{eqnarray} \label{b1}
b_{\gamma} =  \frac{1}{\lambda( \alpha_{opt})}\left(\sqrt{\log (4\gamma)}+\sqrt{K \exp\{-\lambda(\alpha_{opt}) d_{opt}\}}\right)^2,
\end{eqnarray}
will guarantee that our proposed scheme $N^{(soft)}_{\alpha_{opt}}(b_{\gamma},d_{opt})$  satisfies the global false alarm constraint $\gamma$ in the idealized model as in (\ref{equ:constr2}).

Note that all these choices of parameters do not depend on the $\epsilon$ or $g$, and only depend on the idealized model $f_{0}, f_1$ and a prior knowledge on the number $m$  of affected data streams.

\section{Breakdown point analysis}\label{sec:bd}

In this section, we will investigate the robustness properties of our proposed schemes,  $N^{(soft)}_{\alpha}(b,d)$ in (\ref{soft}) and   $N^{(r)}_{\alpha}(b)$ in (\ref{rankCUSUM}), through the false alarm breakdown point analysis. This will provide the guideline on how to choose the tuning parameter $\alpha,$ which controls the robustness of our proposed schemes.

In the classical offline robust statistics, the breakdown point is one of the most popular measures of robustness of statistical procedures. At a high-level,  in the context of finite samples,  the breakdown point is the smallest percentage of contaminations that may cause an estimator or statistical test to be really poor. Since the pioneering work of \cite{hampel:1968} for the asymptotic definition of breakdown point, much research has been done to investigate the breakdown point for different robust estimators or hypothesis testings in the offline statistics, see \cite{krasker:1982}, \cite{Rousseeuw:1984}. To the best of our knowledge, no research has been done on the breakdown point analysis under the online monitoring or change-point context.



Given the importance of the system-wise false alarm rate for online monitoring large-scale data streams in real-world applications, here we focus on the breakdown point analysis for false alarms. Intuitively, for a family of procedures $T(b)$ that is robust, if it is designed to satisfy the false alarm constraint $\gamma$ in (\ref{equ:constr2}) under the idealized model $f_{0},$ then its false alarm rate should not be too bad  under the gross error model $h_{0}$ with some small amount of outliers. There are two specific technical issues that require further clarification. First, how bad is a ``bad" false alarm rate? We propose to follow the sequential change-point detection literature to assess the false alarm rate by $\log \E^{(\infty)}_{h_0}(T(b))$ and deem  the false alarm rate unacceptable if  $\log \E^{(\infty)}_{h_0}(T(b))$ is much smaller than the designed level of $\log\gamma$, i.e., if $\log \E^{(\infty)}_{h_0}(T(b)) = o(\log\gamma).$ Second, what kind of the contamination function $g$ in (\ref{e-neigh}) should we consider in the gross error model?  Here we propose to follow the offline robust statistics literature to consider the worst-case scenario in the $\epsilon$-contaminated distribution class in \cite{huber:1964} that includes any arbitrary  contamination functions $g$'s, which leads to the definition of the false alarm breakdown point in (\ref{breakdownl}).


%

Now we are ready to conduct the false alarm breakdown point analysis for our proposed schemes $N^{(soft)}_{\alpha}(b,d)$ and $N^{(r)}_{\alpha}(b)$ with a given tuning parameter $\alpha \ge 0$. To do so, for the densities $f_{0}(x)$ and $f_{1}(x),$ and for any given $\alpha \ge 0,$   we define an intrinsic bound
\begin{eqnarray} \label{eqnM0M1}
M(\alpha)=\underset{x}{{\rm ess} \sup } \frac{[f_{1}(x)]^{\alpha}-[f_{0}(x)]^{\alpha}}{\alpha},
\end{eqnarray}
and  the density power divergence between $f_{0}$ and $f_{1}$:
 \begin{eqnarray} \label{eqndensitypower}
d_{\alpha}(f_0,f_1)&=&\int\Big\{[f_{1}(x)]^{1+\alpha}-(1+\frac{1}{\alpha})f_{0}(x)[f_{1}(x)]^{\alpha}+\frac{1}{\alpha}[f_{0}(x)]^{1+\alpha}\Big\}d x.
\end{eqnarray}
Note that $d_{\alpha}(f_{0},f_{1})$ was proposed in \cite{basu:1998}, which showed that it is always positive when $f_{1}$ and $f_{0}$  are different. Moreover, when $\alpha = 0,$  $d_{\alpha=0}(f_{0},f_{1})$ becomes Kullback-Leibler information number $\int f_{0}(x)\log \frac{f_{0}(x)}{f_{1}(x)} dx.$

With these two new notations, the following theorem derives the false alarm breakdown point of our proposed schemes $N^{(soft)}_{\alpha}(b,d)$ and $N^{(r)}_{\alpha}(b)$ as a function of the tuning parameter $\alpha$ for a fixed soft-thresholding parameter $d$ and $r$ when online monitoring a given $K$ number of data streams. Since they have the same breakdown point, to simplify the notation, we use $N_{\alpha}$ to denote both the scheme $N^{(soft)}_{\alpha}(b,d)$ and scheme $N^{(r)}_{\alpha}(b).$

\begin{theorem}\label{thm6}
Suppose that $f_{\theta}(x)=f(x-\theta)$ is a location family of density function with continuous probability density function $f(x),$ and assume $f_{\theta_0}(x)-f_{\theta_1}(x)$ takes both positive and negative values for $x\in(-\infty,+\infty)$.
For $\alpha \ge 0,$  and any fixed $d$ and $K,$ the false alarm breakdown point of  our proposed schemes $N_{\alpha}$ in (\ref{soft}) and (\ref{rankCUSUM})  is the same and given by
 \begin{eqnarray} \label{eqnBreakpoint1}
\epsilon^{*}(N_{\alpha})=\frac{d_{\alpha}(f_{\theta_0},f_{\theta_1})}{d_{\alpha}(f_{\theta_0},f_{\theta_1})+(1+\alpha)M(\alpha)},
\end{eqnarray}
where $M(\alpha)$  and $d_{\alpha}(f_{\theta_0},f_{\theta_1})$  are defined in (\ref{eqnM0M1}) and (\ref{eqndensitypower}). In particular, $\epsilon^{*}(N_{\alpha}) = 0$ if $M(\alpha)=\infty$ and $d_{\alpha}(f_{\theta_0},f_{\theta_1})$ is finite.
\end{theorem}

The proof of Theorem \ref{thm6} requires the asymptotic properties of our proposed schemes $N^{(soft)}_{\alpha}(b,d)$  in (\ref{soft}) and $N^{(r)}_{\alpha}(b)$ in (\ref{rankCUSUM}) under the assumption that $\epsilon$ and $g$ are given, which has been studied in the previous  section.
The detailed proof of Theorem \ref{thm6} will be presented in the supplementary materials.

Next, let us apply Theorem \ref{thm6}  to guide us to choose the optimal robustness parameter $\alpha$. Since the false alarm breakdown point of our proposed schemes do not require any information  about the contamination ratio $\epsilon$ and contamination distribution $g,$ one nature idea is to maximize the false alarm breakdown point in (\ref{eqnBreakpoint1}):
\begin{eqnarray}\label{equ:alphaopt2}
\alpha_{opt} = \arg\max_{\alpha \ge 0}\frac{d_{\alpha}(f_{\theta_0},f_{\theta_1})}{d_{\alpha}(f_{\theta_0},f_{\theta_1})+(1+\alpha)M(\alpha)}
\end{eqnarray}

As an illustration, let us see the results of (\ref{eqnBreakpoint1}) and (\ref{equ:alphaopt2}) for widely used normal distributions, i.e., when $f_\theta$ is the pdf of $N(\theta,\sigma^2).$ In this case,  when $\alpha = 0,$ the density power divergence $d_{\alpha=0}(f_{\theta_0},f_{\theta_1}) = \frac{1}{2\sigma^2}(\theta_1-\theta_0)^2$ is finite, but the bound $M(\alpha=0)$  in (\ref{eqnM0M1})  becomes $+\infty$ since it is the supremum of the log-likelihood ratio $\log f_{\theta_1}(x) - \log f_{\theta_0}(x)= (\theta_1-\theta_0) x - (\theta_1^2-\theta_0^2)/2$ over $x \in (-\infty, \infty).$ Hence, $\epsilon^{*}(N_{\alpha=0})=0.$
That is, the false alarm breakdown point of the baseline CUSUM-based scheme $N_{\alpha=0}$
 is $0,$ i.e., any amount of outliers will deteriorate the false alarm rate of the classical CUSUM statistics-based schemes. This is consistent with the offline robust statistics literature that the likelihood-function based methods are very sensitive to model assumptions and are generally not robust.

Meanwhile, for any $\alpha > 0,$ note that
\begin{eqnarray*}
\int_{-\infty}^{\infty} f_{\theta_0}(x) [f_{\theta_1}(x)]^{\alpha} d x  =   \frac{1}{(\sqrt{2\pi}\sigma)^{\alpha} \sqrt{1+\alpha}} \exp\left( -\frac{\alpha (\theta_1-\theta_0)^2}{2(1+\alpha)\sigma^2} \right),
\end{eqnarray*}
and thus it is not difficult to derive  from (\ref{eqndensitypower}) that
\begin{eqnarray}\label{dvalue}
d_{\alpha}(f_{\theta_0},f_{\theta_1})=\frac{\sqrt{1+\alpha}}{\alpha(\sqrt{2\pi} \sigma)^{\alpha}}\left(1- \exp( -\frac{\alpha (\theta_1-\theta_0)^2}{2(1+\alpha)\sigma^2}) \right).
\end{eqnarray}
Moreover, if we let $M (=1/\sqrt{2\pi \sigma^2}),$  then $|f_\theta(x)| \le M$ for all $x.$ By the definition in (\ref{eqnM0M1}), we have $|M(\alpha)| \le 2 M^{\alpha} / \alpha,$ which is finite for any $\alpha > 0.$ This implies that for normal distributions, $\epsilon^{*}(N_{\alpha}) > 0$ for any $\alpha > 0.$ Thus our proposed $L_{\alpha}$-CUSUM based scheme with $\alpha > 0$ is much more robust than the classical CUSUM scheme.

%
%
%

To see the optimal choice of $\alpha$ based on (\ref{equ:alphaopt2}), let us consider a concrete numerical example when $f_{\theta_0} \sim N(0,1)$ and $f_{\theta_1} \sim N(1,1).$  By (\ref{dvalue}),  we can compute the value $d_{\alpha}(0,1)$ for any $\alpha\ge 0.$ While we do not have analytic formula for the upper bound $M(\alpha)$ in (\ref{eqnM0M1}), its numerical value can be easily found by brute-force exhaustive search over the real line $x \in (-\infty, \infty).$ The result shows the false alarm breakdown point of our proposed scheme $N_{\alpha}$ will first increase and then decrease as $\alpha$ varies from $0$ to $2.$, and yields the optimal choice of $\alpha_{opt}$ as $0.51,$ with corresponding breakdown point as $0.233.$ That means our proposed scheme with the choice of $\alpha=0.51$ could tolerate $23.3\%$ arbitrarily bad observations in terms of keeping the designed false alarm constraint stable.

\begin{figure}
  \begin{minipage}[t]{0.49\linewidth}
  \includegraphics[width=1\linewidth]{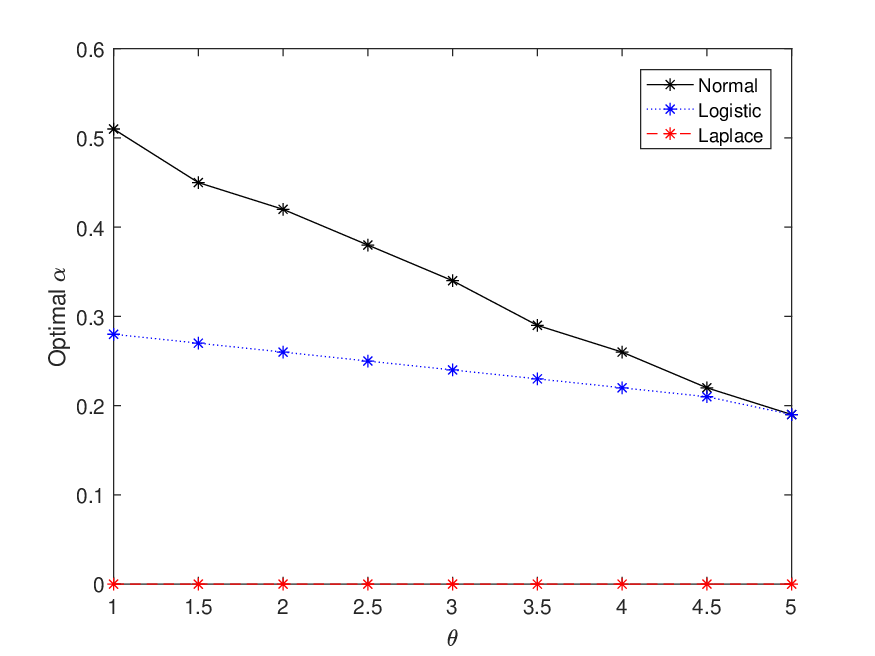}
   \caption{The value $\alpha_{opt}$ in (\ref{equ:alphaopt2}) for different $\theta_1.$}\label{fig:optalpha}
  \end{minipage}
    \hfill
\begin{minipage}[t]{0.49\linewidth}
\includegraphics[width=1\linewidth]{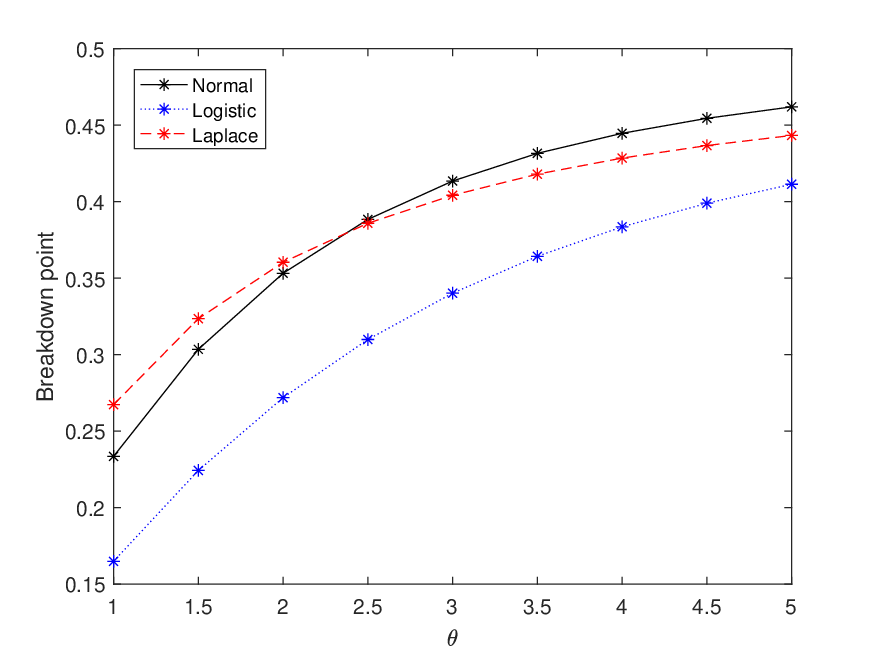}
    \caption{The false alarm breakdown point $\epsilon^{*}(N_{\alpha})$ in (\ref{eqnBreakpoint1}) when $\alpha = \alpha_{opt}$ for different $\theta_1.$}
    \label{fig:optbreakdown}
\end{minipage}
\end{figure}

Finally, we should emphasize that the optimal value $\alpha_{opt}$ in (\ref{equ:alphaopt2}) and the false alarm breakdown point  $\epsilon^{*}(N_{\alpha})$ in (\ref{eqnBreakpoint1}) will generally depend on the change magnitude or signal-to-noise-ratio. To illustrate this, we consider three families: Normal, Laplace, and Logistic distributions with the scale parameter $\sigma=1.$ This yields three families of pdfs, $f_{\theta}(x)=\frac{1}{\sqrt{2\pi}}\exp(-\frac{(x-\theta)^2}{2}),$ $\frac{1}{2}\exp(-|x-\theta|)$, or $\frac{\exp(-(x-\theta))}{(1+\exp(-(x-\theta)))^2}.$  In each case, we assume that the pre-change parameter $\theta_0=0,$  the designed post-change parameter $\theta_1$ varies from $1$ to $5.$  In Figures \ref{fig:optalpha} and  \ref{fig:optbreakdown}, we plot the optimal value $\alpha_{opt}$ and the corresponding false alarm breakdown point  $\epsilon^{*}(N_{\alpha})$  as a function of $\theta_1.$ 
Figure \ref{fig:optbreakdown} implies that  with the increasing of the post-change $\theta_1$ or the signal-to-noise-ratio, our proposed robust schemes with optimal $\alpha$ can tolerate more outliers. Also it is interesting to see from Figure \ref{fig:optalpha} that the optimal $\alpha_{opt}$ decreases for normal or logistic distribution as the post-change parameter $\theta_1$ increases. A surprising result is that the optimal $\alpha_{opt} = 0$  for the Laplace distribution. This implies the classical CUSUM procedure for Laplace distribution is actually optimal in the sense of having the largest breakdown point. One possible explanation is that for the Laplace distribution, the log-likelihood ratio $\log (f_{\theta_1}(x)/f_{\theta_0}(x))= -|x-\theta_1| + |x-\theta_0|$  takes values in the interval $[\theta_0-\theta_1, \theta_1 -\theta_0]$ when $\theta_1 > \theta_0.$ Thus the impact of outliers is directly controlled.

\section{Numerical Simulations}\label{sec:simu}

In this section we conduct numerical simulation studies to illustrate the robustness and
efficiency of our proposed schemes  $N^{(soft)}_{\alpha}(b,d)$  and $N^{(r)}_{\alpha}(b)$.

In our simulation studies, we assume there are $K= 100$ independent data streams, and at some unknown time, $m=10$ features are affected by the occurring event. Also the change is instantaneous if a stream is affected, and we do not know which subset of streams will be affected. We set $f_{\theta}=$ pdf of $N(\theta,1),$ the pre-change parameter $\theta_0=0,$ the post-change parameter  $\theta_1=1,$ and the contamination densities $g_0, g_1$ are pdfs of $N(0, 3^2).$ Our proposed schemes  $N^{(soft)}_{\alpha}(b,d)$   in (\ref{soft}) and  $N^{(r)}_{\alpha}(b)$ in (\ref{rankCUSUM}) are constructed by using the density function $f_{\theta_0}$ and $f_{\theta_1}.$

In the first simulation study, we consider the idealized model when $\epsilon=0.$ In this case, for our proposed robust scheme  $N^{(soft)}_{\alpha}(b,d)$   in (\ref{soft}), as shown in the previous section, the optimal choices of $\alpha_{opt}=0.51.$  By (\ref{d1}), if $\log(\gamma) << K,$ then the corresponding optimal shrinkage parameters $d \approx \frac{1}{\lambda(\epsilon=0, \alpha=0.51)} \log\frac{K}{m} = 0.8915$  for $K= 100$ and $m=10,$ since $\lambda(\epsilon=0,\alpha=0.51)=2.5829.$  For our proposed robust scheme  $N^{(r)}_{\alpha}(b)$ in (\ref{rankCUSUM}), we choose $\alpha=\alpha_{opt}=0.51$ and $r=10.$
For the baseline CUSUM-based scheme, i.e., $N^{(soft)}_{\alpha=0}(b,d)$ with $\alpha = 0$, we choose the shrinkage parameter $d=\frac{1}{\lambda(\epsilon=0, \alpha=0)} \log\frac{K}{m} = 2.3026,$ since $\lambda(\epsilon=0, \alpha=0)=1.$

In summary,  we will  compare  the following different schemes.

\begin{itemize}
  \item  Our proposed scheme $N^{(soft)}_{\alpha}(b, d)$   in (\ref{soft}) with $\alpha_{opt} = 0.51$ and $d=0.8915.$
    \item  Our proposed scheme $N^{(r)}_{\alpha}(b)$   in (\ref{rankCUSUM}) with $\alpha_{opt} = 0.51$ and $r=10.$
  \item  The baseline CUSUM-based scheme $N^{(soft)}_{\alpha=0}(b,d)$ with $d=2.3026.$
    \item  The MAX scheme $N_{\alpha = 0.51,\max}(b)$ in    (\ref{eqnmax});
  \item  The SUM scheme $N_{\alpha = 0.51,\text{sum}}(b)$  in  (\ref{eqnsum});
  \item  The method $N_{XS}(b,p_0=0.1)$ in \cite{xie:2013} based on generalized likelihood ratio:
  \begin{eqnarray*} \label{eqn_S}
 N_{XS}(b, p_0)=
\inf\Big\{n \ge 1: \max_{0\le i < n} \sum_{k=1}^{K} \log(1-p_0+&p_0 \exp[ \big(U_{k,n,i}^{+}\big)^2/2]) \ge b \Big\},
\end{eqnarray*}
where  for all $1 \le k \le K, 0 \le i < n,$
\[
U_{k,n,i}^{+}= \max\big(0, \frac{1}{\sqrt{n-i}} \sum_{j=i+1}^n X_{k,j}\big).
\]
 \item  The method $N_{Chan,\alpha=0}(b,p_0=0.1)$ in \cite{chan:2015} under the idealized model that is an extension of the SUM scheme in  \cite{mei:2010}:
     \begin{eqnarray*} \label{eqn_Chan2}
N_{Chan,\alpha=0}(b,p_0) =\inf\Big\{n \ge 1:  \sum_{k=1}^{K} \log \big(1-p_0+0.64*p_0 \exp(W_{k,n}^{*}/2)\big)  \ge b \Big\},
\end{eqnarray*}
where $W_{k,n}^{*}$ is the CUSUM statistics in (\ref{eq3n011}).
\item  The method $N_{Chan,\alpha=0.51}(b,p_0=0.1)$ in (\ref{eqn_Chana}) which is similar to $N_{Chan,\alpha=0}$ but replace the CUSUM statistic by our proposed $L_{\alpha}$-CUSUM statistic.
\end{itemize}


For each of these schemes $T(b)$,  we first find the appropriate values of the  threshold $b$ to satisfy the false alarm constraint $\gamma \approx 5000$ under the idealized model with $\epsilon=0$ (within the range of sampling error). Next, using the obtained global threshold value $b,$ we simulate the detection delay when the change-point occurs at time $\nu=1$ under several different post-change scenarios, i.e., different number of affected sensors. All Monte Carlo simulations are based on $1000$  repetitions.

\begin{table}[h]
 \centering
  \caption{A comparison of the detection delays of $8$ schemes with $\gamma = 5000$ under the idealized model.
  The smallest and largest standard errors of these $8$ schemes are also reported under each post-change hypothesis
  based on $1000$ repetitions in Monte Carlo simulations.
  }\label{table05}
  \smallskip

  \begin{tabular}{|l|c | c|c|c|c|c|c|c|}
    \hline
  \multicolumn{9}{|c|}{Gross error model with $\epsilon=0$} \\
   \hline
  &\multicolumn{8}{|c|}{\# affected local data streams} \\
\cline{2-9}
         & $1$& $3$& $8$& $10$&$15$& $20$ &$50$ & $100$ \\ \hline
 \hline
          Smallest standard error            & $0.29$ & $0.12$  & $0.05$& $0.04$ &$0.03$& $0.03$ &$0.01$ & $0.00$ \\ \hline
        Largest standard error            & $0.58$ & $0.20$ &   $0.07$& $0.06$ &$0.05$& $0.03$&$0.02$ & $0.01$  \\ \hline
      \hline
               \multicolumn{9}{|c|}{Our proposed robust scheme } \\ \hline
 $N^{(soft)}_{\alpha=0.51}(b=8.5,d=0.8915)$     & $41.0$ & $18.6$  &$10.3$ &$9.2$&$7.5$ &$6.5$ &$4.5$ &$3.9$  \\ \hline
      $N^{(r=10)}_{\alpha=0.51}(b=17.19)$     & $40.6$ & $18.5$ & $10.3$ &$9.2$&$7.7$ &$6.9$  &$5.3$ &$4.8$  \\ \hline    \hline
   \multicolumn{9}{|c|}{Comparison of other methods } \\ \hline
      $N^{(soft)}_{\alpha=0}(b=21.52,d=2.3026)$        & $33.6$ & $15.2$ & $8.4$& $7.5$&$6.1$ &$5.3$  &$3.7$ & $3.0$ \\  \hline
    $N_{\alpha=0.51, \max}(b=4.3)$         & $27.7$ & $19.6$ & $16.2$ & $15.6$ &$14.8$ &$14.2$ &$12.7$ & $11.9$ \\  \hline
 $N_{\alpha=0.51, \text{sum}}(b=36.85)$        & $63.7$ & $26.9$  &$12.5$ & $10.5$ &$7.8$ &$6.4$   &$3.3$ & $2.0$ \\  \hline
           $N_{Chan,\alpha=0.51}(b=1.04,p_0=0.1)$        & $31.4$ &$17.7$&$10.8$&$9.7$&$7.8$&$6.7$&  $4.1$& $3.0$ \\  \hline
   $N_{Chan,\alpha=0}(b=21.6,p_0=0.1)$        & $32$ &$15.2$&$11.2$&$7.5$&$5.3$&$4.2$& $3.3$& $2.3$ \\  \hline
          $N_{XS}(b=19.5, p_0=0.1)$        & $30.9$ & $13.2$ &  $7.2$ &$5.7$  & $4.7$ &$3.5$ &$1.8$&$1.0$ \\  \hline
  \end{tabular}

\end{table}

\begin{table}[h]
 \centering

  \caption{A comparison of the detection efficiency score of $8$ schemes under the gross error model with $\epsilon=0.1$ based on $1000$ repetitions in Monte Carlo simulations. The threshold $b$ is chosen to satisfy $\gamma=5000$ in the idealized model.
  }\label{table01}
  \smallskip
 \begin{tabular}{|l|c | c|c|c|c|c|c|c|}    \hline
  \multicolumn{9}{|c|}{Gross error model with $\epsilon=0.1$} \\
   \hline
  &\multicolumn{8}{|c|}{\# affected local data streams} \\
\cline{2-9}
         & $1$& $3$ &$8$& $10$&$15$& $20$ & $50$ & $100$ \\ \hline
 \hline
               \multicolumn{9}{|c|}{Our proposed robust scheme } \\ \hline
 $N^{(soft)}_{\alpha=0.51}(b=8.5,d=0.8915)$     & $0.17$ & $0.34$ & $0.61$ &$0.68$&$0.83$ &$0.95$  &$1.37$ &$1.66$  \\ \hline
   $N^{(r=10)}_{\alpha=0.51}(b=17.09)$     & $0.17$ & $0.35$ & $0.62$ &$0.68$&$0.82$ &$0.92$  &$1.2$ &$1.35$  \\ \hline    \hline
   \multicolumn{9}{|c|}{ Other methods for comparison } \\ \hline
         $N^{(soft)}_{\alpha=0}(b=21.52,d=2.3026)$        & $0.27$ & $0.32$ & $0.4$& $0.43$&$0.48$ &$0.53$  &$0.7$ & $0.8$ \\  \hline
    $N_{\alpha=0.51, \max}(b=4.3)$         & $0.3$ & $0.36$ & $0.44$ & $0.46$ &$0.49$ &$0.51$ &$0.58$ & $0.62$ \\  \hline
   $N_{\alpha=0.51, \text{sum}}(b=36.85)$        & $0.12$ & $0.25$ & $0.5$ & $0.58$ &$0.77$ &$0.94$   &$1.73$ & $2.88$ \\  \hline
    $N_{Chan,\alpha=0}(b=21.6,p_0=0.1)$        & $0.26$ &$0.32$&$0.36$&$0.43$&$0.54$&$0.63$&  $0.75$& $1.03$ \\  \hline
        $N_{Chan,\alpha=0.51}(b=1.04,p_0=0.1)$        & $0.21$ &$0.37$&$0.59$&$0.65$&$0.81$&$0.94$&  $1.53$& $2.19$ \\  \hline
$N_{XS}(b=19.5, p_0=0.1)$        & $0.22$ & $0.39$ &  $0.44$ &$0.49$  & $0.52$ &$0.55$ &$0.78$&$0.93$ \\ \hline
  \end{tabular}

\end{table}

Table \ref{table05} summarizes the detection delays of these $8$ schemes under $9$ different post-change hypothesis. Among all schemes,  $N_{XS}(b,p_0)$  generally yields the smallest detection delay. However, we want to emphasize that it is  computationally  expensive. Specifically, even if we use a time window of size $k$ as in \cite{xie:2013} to speed up the implementation of $N_{XS}(b,p_0)$, at each time $n,$ $O(K k^2)$ computations are needed to get the global monitoring statistics, whereas our proposed scheme $N^{(soft)}_{\alpha}(b, d)$  only require $O(K)$ computations to get the global monitoring statistics.

 Another interesting observation from Table \ref{table05} is that the detection delay of our proposed robust schemes $N^{(soft)}_{\alpha=0.51}(b, d)$ and $N^{(r)}_{\alpha=0.51}(b)$  are not too bad compared with the CUSUM-based scheme $N^{(soft)}_{\alpha=0}(b,d=2.3026)$, and it just takes additional $1.7$ time steps to raise a correct global alarm under the idealized model when $m=10$ data streams are affected.

In the second simulation study, we will examine the detection efficiency of these schemes under the gross error model when $\epsilon=0.1.$ For each of these $8$ schemes, we use the same threshold $b$ obtained from the first simulation to guarantee these schemes satisfy the same false alarm constraint $\gamma=5000$ under the idealized model. Then, we wilsimulate the in-control average run and the detection delay of these schemes when both the pre-change distribution and post-change distribution are the gross error model in  (\ref{equ:highgrosserror}) with $\epsilon=0.1,$ $g_0,g_1$ as pdfs of $N(0,3^2).$ We then report the empirical version of the asympototic efficiency score in (\ref{eq:ae}) of these schemes under $8$ different post-change hypothesis in Table \ref{table01}.

First, we can see our proposed scheme $N^{(soft)}_{\alpha=0.51}(b,d=0.8915)$ and $N^{(r=10)}_{\alpha=0.51}(b=18.7)$ have the largest detection efficiency score among all comparison methods when $10$ data streams are affected. Moreover, by using our proposed $L_{\alpha}$-CUSUM statistics with $\alpha = 0.51,$ the method $N_{Chan,\alpha=0.51}(b,p_0=0.1)$ yields the similar detection efficiency to our proposed schemes.  This illustrates that the improvement of $L_{\alpha}$-CUSUM statistics is significant as compared to the baseline CUSUM  statistics in the presence of outliers.



It is also interesting to note that the MAX-scheme $N_{\alpha=0.51,\max}(b)$ and the SUM-scheme $N_{\alpha=0.51,\text{sum}}(b)$ are designed for the case when $m=1$ or $m=K$ features are affected, and Table \ref{table01} confirmed that their detection efficinecys are indeed the largest in their respective designed scenarios. However, when the number of affected features $m$ is moderate and is arround $10$, our proposed scheme $N^{(soft)}_{\alpha=0.51}(b,d)$ and  $N^{(r)}_{\alpha=0.51}(b)$  have larger detection efficiency, which implies our proposed schemes with sum-shrinkage technique could be more robust to the number of affected features.

In the third experiment, we investigate the impact of contamination rate $\epsilon$ on the false alarms of different methods to illustrate the robustness of our proposed $L_{\alpha}$-CUSUM statistics. Since the top-r scheme $N^{(r)}_{\alpha=0.51}(b),$ MAX-scheme  $N_{\alpha=0.51, \max}(b),$ the SUM-scheme $N_{\alpha=0.51, \text{sum}}(b)$ and $N_{Chan,\alpha=0.51}(b,p_0)$  are all based on local $L_{\alpha}$-CUSUM statistics,
their robustness properties are similar to our proposed scheme $N^{(soft)}_{\alpha=0.51}(b,d).$
To highlight the robustness of our proposed $L_{\alpha}$-CUSUM statistics,  we only compare our proposed scheme $N^{(soft)}_{\alpha=0.51}(b,d)$  with other three schemes: $N^{(soft)}_{\alpha=0}(b,d)$, $N_{Chan,\alpha=0}(b,p_0),$ and $N_{XS}(b, p_0).$



\begin{figure}[!t]
        \centering
  \includegraphics[height=2.2in]{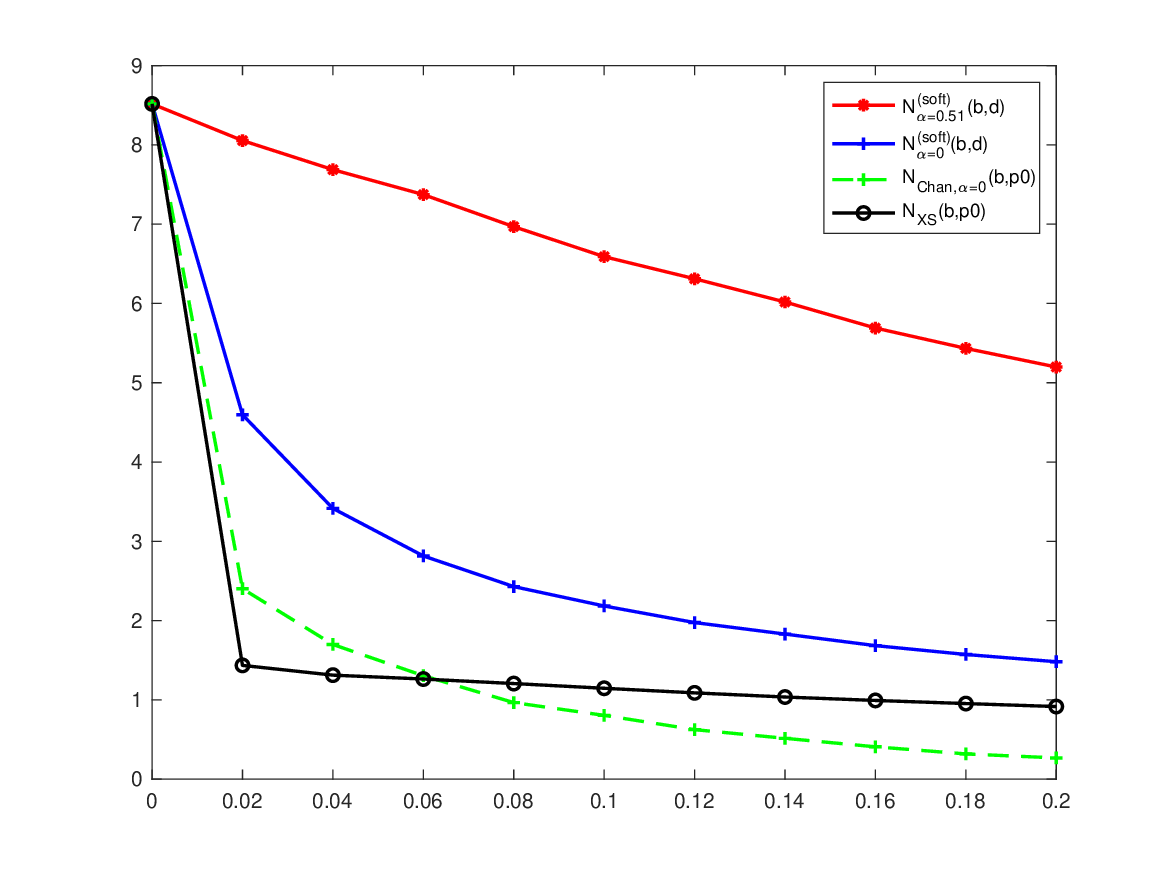}
 \caption{Each line represents $\log \E_{h_{0}}^{(\infty)}(T)$ of a scheme as a function of $\epsilon \in (0, 0.2).$ } \label{fig:icarl}
\end{figure}

Figure \ref{fig:icarl} reports the curve of $\log \E_{h_{0}}^{(\infty)}(T)$  as the contamination ratio $\epsilon$ varies from $0.02$ to $0.2$ with stepsize $0.02$. Clearly, all curves decrease with the increasing of contaminations, meaning that all schemes will raise false alarm more frequently when there are more outliers.
However, the curves for the CUSUM or likelihood-ratio based methods decreased very quickly, whereas our proposed $L_{\alpha}$-CUSUM statistics-based method with $\alpha_{opt}=0.51$ decreases rather slowly. This suggests that our proposed scheme is more robust in the sense of keeping the designed ARL more stable with a small departure from the assumed model.

\section{Conclusion}\label{sec:con}
In this paper, we study the problem of robust monitoring of large-scale data streams when the true observed data follow Huber's gross error model. We develop a family of efficient and robust detection schemes that can be implemented in real-time. From the worst-case detection efficiency point of view, we show our proposed methods can still have positive detection efficiency under a small proportion of arbitrary outliers. In contrast, the CUSUM-based methods lose all detection efficiency once the data include outliers. From the robustness point of view, we propose a new concept called false alarm breakdown point, which measures the stability of the designed false alarm constraint of any monitoring procedures under the effects of outliers. Our breakdown point analysis implies our proposed methods can have positive breakdown points.  We also provide detailed guidelines on the choices of tuning parameters in our detection procedures. However, in this work, we focus on the problem of monitoring homogeneous independent data streams. It  is  of future  interest  to  extend the problem to nonhomogeneous data streams with some correlation structures.

\section*{Appendix}

In this online supplementary material, we provide detailed proofs to Theorems \ref{thm1}, \ref{thm1b}, and Theorem \ref{the:wae},  the optimal parameter choice in Section \ref{sec:dae}, and the proof of  Theorem \ref{thm6}.

\subsection*{A. Proof of Theorem \ref{thm1}} (a)  For any $x \ge 0,$ by Chebyshev's inequality,
\begin{eqnarray} \label{eqnproofthm01a}
\E_{h_0}^{(\infty)}[N^{(soft)}_{\alpha}(b,d)] &\ge& x \Prob_{h_0}^{(\infty)}(N^{(soft)}_{\alpha}(b,d) \ge x) \cr
&=& x \left[1 -  \Prob_{h_0}^{(\infty)}(N^{(soft)}_{\alpha}(b,d) < x) \right] \cr
&=& x \left[1 -  \Prob_{h_0}^{(\infty)}( \sum_{k=1}^{K} \max\{0, W_{\alpha, k,n}-d\} \ge b) \mbox{ for some } 1 \le n \le x \right] \cr
&\ge&   x \left[1 -  x \Prob_{h_0}^{(\infty)}( \sum_{k=1}^{K} \max\{0, W_{\alpha, k}^{*}-d\} \ge b)\right],
\end{eqnarray}
where $W_{\alpha, k}^{*} = \limsup_{n \rightarrow \infty} W_{\alpha, k, n}.$ We will show that $W_{\alpha, k}^{*}$ exists later, and when it does exist, it is clear that $W_{\alpha, k}^{*}$ are i.i.d. across different $k$ under the pre-change measure $\Prob_{h_0}^{(\infty)}.$ Now if we define  the log-moment generating function of the $W_{\alpha, k}^{*}$'s
\begin{eqnarray} \label{eqnMGF}
\psi_{\alpha}(\theta) = \log \E^{(\infty)}_{h_0}\exp\{\theta \max(0,W_{\alpha,k}^{*}-d)\}
\end{eqnarray}
for some $\theta \ge 0,$ then another round application of Chebyshev's inequality yields
\begin{eqnarray} \label{eqnproofthm01b}
\exp(K \psi_{\alpha}(\theta)) &=& \E^{(\infty)}_{h_0}\exp\{\theta \sum_{k=1}^{K} \max(0,W_{\alpha,k}^{*}-d)\}  \cr
&\ge& e^{\theta b}  \Prob_{h_0}^{(\infty)}( \sum_{k=1}^{K} \max\{0, W_{\alpha, k}^{*}-d\} \ge b)
\end{eqnarray}
for $\theta > 0.$ Combining (\ref{eqnproofthm01a}) and (\ref{eqnproofthm01b})  yields that
\begin{eqnarray} \label{eqnproofthm01c}
\E_{h_0}^{(\infty)}[N^{(soft)}_{\alpha}(b,d)] &\ge& x \left[1 -  x \exp(-\theta b + K \psi_{\alpha}(\theta)) \right]
\end{eqnarray}
for all $x \ge 0.$ Since $x(1- xu)$ is maximized at $x=1/(2u)$ with the maximum value $1/(4u).$ We conclude from (\ref{eqnproofthm01c}) that
\begin{eqnarray} \label{eqnproofthm01d}
\E_{h_0}^{(\infty)}[N^{(soft)}_{\alpha}(b,d)] &\ge& \frac14 \exp\left(\theta b - K \psi_{\alpha}(\theta)\right).
\end{eqnarray}
for any $\theta > 0$ as long as $\psi_{\alpha}(\theta)$ in (\ref{eqnMGF}) is well-defined.

The remaining proof is to utilize the definition of $\lambda(\epsilon, \alpha; g_0) > 0$ in (\ref{constantlambda})   to show that the upper limiting $W_{\alpha, k}^{*}$ of the proposed $L_{\alpha}$-CUSUM statistics is well-defined and derive a careful analysis of  $\psi_{\alpha}(\theta)$ in (\ref{eqnMGF}).
When $\alpha = 0,$ the $L_{\alpha}$-CUSUM statistics become the classical CUSUM statistics, and the corresponding analysis is well-known, see \cite{liu:2017}. Here our main insight is that our proposed $L_{\alpha}$-CUSUM statistics $W_{\alpha,k, n}$ for detecting a change from $h_{0}(x)$ to $h_{1}(x)$ in (\ref{equ:highgrosserror}) can be thought of as the classical CUSUM statistic for detecting a local change from $h_{0}(x)$ to another new density function $h_{2}(x).$ Hence, under the pre-change hypothesis of $h_0(\cdot),$ the false alarm properties of  our proposed $L_{\alpha}$-CUSUM statistics can be derived through those of the classical CUSUM statistics.

By the definition of $\lambda(\epsilon,\alpha;g_0) > 0$, if we define a new function
\begin{eqnarray}\label{h2}
h_{2}(x):= \exp\left\{\lambda(\epsilon,\alpha;g_0)(\frac{(f_{1}(x))^{\alpha}-(f_{0}(x))^{\alpha}}{\alpha})\right\}h_{0}(x),
\end{eqnarray}
then  $h_2(x)$ is a well-defined probability density function. Then in the problem of detection a local change from $h_{0}(x)$ to $h_{2}(x),$ the local CUSUM statistics for the $k$th local data stream is defined recursively by
\begin{eqnarray*}
W_{k, n}' &=& \max\{0,W_{k, n-1}'+\log\frac{h_{2}(X_{k,n})}{h_{0}(X_{k,n})}\}\nonumber \\
&=&\max\{0,W_{k,n-1}'+\lambda(\epsilon,\alpha;g_0)\frac{[f_{1}(X_{k,n})]^{\alpha}-[f_{0}(X_{k,n})]^{\alpha}}{\alpha}\}.
\end{eqnarray*}
Compared with our proposed $L_{\alpha}$-CUSUM statistics $W_{\alpha,k,n}$, it is clear that $W_{k, n}'=\lambda(\epsilon,\alpha) W_{\alpha,k,n},$
and thus our proposed $L_{\alpha}$-CUSUM statistics $W_{\alpha,k,n}$'s are equivalent to the standard CUSUM statistics $W_{k, n}'$ up to a positive constant $\lambda(\epsilon,\alpha; g_0).$
By the classical results on the CUSUM, see Appendix 2 on Page 245 of  \cite{siegmund:1985},
as $n \rightarrow \infty,$ $W_{k, n}'$ converges to a limit and thus
$W_{\alpha,k,n}$ also converges to a limit, denoted by $W_{\alpha,k}^{*}.$ Moreover, the tail probability of $W_{\alpha,k}^{*}$ satisfies
\begin{eqnarray}\label{expon}
G(x) = \Prob^{(\infty)}_{\theta_0}(W_{\alpha,k}^* \ge x)  = \Prob^{(\infty)}_{\theta_0}( \limsup_{n \rightarrow \infty} W_{k,n}' \ge \lambda(\epsilon,\alpha; g_0) x)  \le e^{-\lambda(\epsilon,\alpha;g_0)x}.
\end{eqnarray}

Now  we shall use (\ref{expon}) to derive information bound of  $\psi_{\alpha}(\theta)$ in (\ref{eqnMGF}). In order to simplify our arguments, we abuse the notation and simply denote $\lambda(\epsilon,\alpha; g_0)$  by $\lambda$ in the remaining proof of the theorem. By the definition of $\psi_{\alpha,k}(\theta)$ in (\ref{eqnMGF}) and the tail probability $G(x)$ in (\ref{expon}), for $\theta > 0,$
\begin{eqnarray} \label{eqnWasym2}
\psi_{\alpha}(\theta) &=& \log [  \Prob_{\theta_0}^{(\infty)}(W_{\alpha,k}^{*} \le d) - \int_{d}^{\infty} e^{\theta (x-d)}dG(x)] \\
&=& \log [1+ \theta \int_{d}^{\infty} e^{\theta (x-d)}G(x)dx]\nonumber \\
&\le& \log [1+ \theta \int_{d}^{\infty} e^{\theta (x-d)}e^{-\lambda x} dx]\nonumber \\
&=&\log\left(1 + \frac{\theta}{\lambda-\theta}  e^{-d\lambda} \right) \le \frac{\theta}{\lambda-\theta}  e^{-d\lambda},\nonumber
\end{eqnarray}
where the second equation is based on the integration by parts. Clearly, relation (\ref{eqnWasym2}) holds for any $0 < \theta < \lambda= \lambda(\epsilon, \alpha;g_0).$

By (\ref{eqnproofthm01d}) and (\ref{eqnWasym2}),  we have
\begin{eqnarray}  \label{eqnproofthm01f}
\E_{\epsilon}^{\infty}N^{(soft)}_{\alpha}(b,d)&\ge& \frac14  \exp\Big( \theta b- \frac{K\theta }{\lambda-\theta} e^{-d\lambda} \Big)
\end{eqnarray}
for all $0 < \theta < \lambda = \lambda(\epsilon, \alpha; g_0).$ When  $\lambda b > K \exp\{- \lambda d\},$ relation (\ref{a}) follows at once from (\ref{eqnproofthm01f}) by letting $\theta = \sqrt{\lambda/b} \left(\sqrt{\lambda b} - \sqrt{K \exp\{-d\lambda\}}\right) \in (0, \lambda).$ This completes the proof of Theorem \ref{thm1} (a).

(b) Note $N^{(soft)}_{\alpha}(b,d=0)\le N^{(r)}_{\alpha}(b)$ for any $b\ge 0.$ Therefore, (\ref{aa}) can be derived directly from (\ref{a}) by letting $d=0$ in (\ref{a}).

\subsection*{B. Proof of Theorem  \ref{thm1b}}

First, we will prove the part (a) of Theorem  \ref{thm1b}. To prove the detection delay bound (\ref{bound}) in Theorem  \ref{thm1b}, without loss of generality, assume the first $m$ data streams are affected. Consider  a  new stopping time
\begin{eqnarray*}
T'(b,d) &=& \inf\{n \ge 1: \sum_{k=1}^{m}(W_{\alpha,k,n}-d)\ge b\} = \inf\{n \ge 1: \sum_{k=1}^{m} W_{\alpha,k,n} \ge b + md\}.
\end{eqnarray*}
Clearly  $N^{(soft)}_{\alpha}(b,d) \le T'(b,d),$ and thus
\begin{eqnarray*}
\vv D_{h_1}(N^{(soft)}_{\alpha}(b,d))&\le& \vv D_{h_1}(T'(b,d)).
\end{eqnarray*}
Next, by the recursive definition of $W_{\alpha,k,n}$ in (\ref{eqLalpha}), using the same approach in Theorem 2 of \cite{lorden:1971} that connects the recursive CUSUM-type scheme to the random walks, we have
\begin{eqnarray*}
\vv D_{h_1}(T'(b,d)))&\le& \E_{1} T''(b,d),
\end{eqnarray*}
where $\E_{1}$ denotes the expectation when the change happen at time $\nu=1,$ and $T''(b,d)$ is the first passage time when the random walk with i.i.d. increment of mean  $m I_{1}(\epsilon,\alpha;g_1)$  exceeds the bound $b+md,$ and is defined as
\begin{eqnarray*}
T''(b,d) &=& \inf\{n \ge 1:  \sum_{i=1}^{n} \sum_{k=1}^{m} \frac{[f_{1}(X_{k,i})]^{\alpha}-[f_{0}(X_{k,i})]^{\alpha}}{\alpha} \ge b + md \}.
\end{eqnarray*}
By standard renewal theory, as ($\frac{b}{m}+d$)$\to \infty$, we have
$$
\E_{1} T''(b,d) \le \frac{1 + o(1)}{ m I_{1}(\epsilon,\alpha; g_1)} \left( b + m d \right).
$$
Relation (\ref{bound}) then follows at once from  the above relations, which completes the proof of  part (a) of Theorem \ref{thm1b}.

To prove the part (b), we define another stopping time
\begin{eqnarray*}
  \tau(b):=\inf\{n \ge 1: \sum_{k=1}^{m}W_{\alpha,k,n}\ge b\}.
\end{eqnarray*}
Note for the sorted statistics $W_{\alpha,(1),n}\ge W_{\alpha,(2),n}\ge\cdots\ge W_{\alpha,(K),n},$ we have $\sum_{k=1}^{m}W_{\alpha,k,n}\le \sum_{k=1}^{m}W_{\alpha,(k),n}.$ Thus,when $m\le r,$  $N^{(r)}_{\alpha}(b)\le  \tau(b).$ By standard renew theory, we have
\begin{eqnarray*}
\vv D_{h_1}(N^{(r)}_{\alpha}(b))&\le&\vv D_{h_1}(\tau(b))\le (1+o(1))\frac{b}{mI_{\theta}(\epsilon,\alpha)},
\end{eqnarray*}
which completes the proof of  part (b) of Theorem \ref{thm1b}.

\subsection*{C. Proof of Theorem \ref{the:wae}}
Note if $\epsilon<-I_0(\alpha)/[M^*(\alpha)-I_0(\alpha)],$
\begin{eqnarray}
\underset{g_0\in \mathcal G}{\sup} I_0(\epsilon,\alpha;g_0)&=&(1-\epsilon)I_0(\alpha)+\epsilon \,\underset{x}{\sup }(\frac{[f_{1}(x)]^{\alpha}-[f_{0}(x)]^{\alpha}}{\alpha})\nonumber\\
&\le&(1-\epsilon)I_0(\alpha)+\epsilon M^*(\alpha)< 0.
\end{eqnarray}
 Therefore,  by Theorem \ref{thm1}, there exists a postive number $\lambda^*(\epsilon,\alpha)=\underset{g_0\in \mathcal G}{\inf}\lambda(\epsilon,\alpha;g_0)>0$ such that
$$
\lim_{b\to\infty}\frac{\underset{g_0\in \mathcal G}{\inf}\left[\log(\E^{(\infty)}_{h_0}(N_{\alpha}(b)))\right]}{b}\ge \lambda^*(\epsilon,\alpha).
$$
Moreover, if $\epsilon<I_1(\alpha)/[M^*(\alpha)+I_1(\alpha)],$
\begin{eqnarray}
\underset{g_1\in \mathcal G}{\inf} I_1(\epsilon,\alpha;g_1)&=&(1-\epsilon)I_1(\alpha)+\epsilon \,\underset{x}{\inf }(\frac{[f_{1}(x)]^{\alpha}-[f_{0}(x)]^{\alpha}}{\alpha})\nonumber\\
&\ge&(1-\epsilon)I_1(\alpha)-\epsilon M^*(\alpha)> 0.
\end{eqnarray}
By Theorem \ref{thm1b}, we have
\begin{eqnarray*}
\lim_{b\to\infty}\frac{\underset{g_1\in \mathcal G}{\sup}\left[\vv D_{h_1}(N_{\alpha}(b))\right]}{b}&\le& \frac{1}{m\underset{g_1\in \mathcal G}{\inf} I_1(\epsilon,\alpha;g_1)}\le \frac{1}{m [(1-\epsilon)I_1(\alpha)-\epsilon M^*(\alpha)]}.
\end{eqnarray*}
Thus, by the definition of worst-case detection efficiency in (\ref{eq:wae}),  we have
\begin{eqnarray}
\text{WAE}(N_{\alpha}, \epsilon)&=&\lim_{b\to\infty}\frac{\underset{g_0\in \mathcal G}{\inf}\left[\log(\E^{(\infty)}_{h_0}(N_{\alpha}(b)))\right]}{\underset{g_1\in \mathcal G}{\sup}\left[\vv D_{h_1}(N_{\alpha}(b))\right] }\ge m\lambda^*(\epsilon,\alpha)\Big[(1-\epsilon)I_1(\alpha)-\epsilon M^*(\alpha)\Big]. \nonumber
\end{eqnarray}

\subsection*{D. Parameter Setting in Section \ref{sec:dae}}
The choice of $b = b_{\gamma}$ in (\ref{b1}) follows directly from Theorem \ref{thm1} (a). To prove (\ref{d1}),
we abuse the notation and use $\lambda$ to denote $\lambda(\alpha)$ for simplification. By Theorem  \ref{thm1b}, the optimal $d$ is the non-negative value that minimize the function
\begin{eqnarray}\label{findd}
\ell(d):=   \frac{b_{\gamma}}{m}+d  = \frac{1}{\lambda m }(\sqrt{\log (4\gamma)}+\sqrt{Ke^{-\lambda d}})^2+d.
\end{eqnarray}
This is an elementary optimization problem, and the optimal $d$ can be found by taking derivative of $\ell(d)$ with respect to $d$, since $\ell(d)$ is a convex function of $d.$ To see this,
\begin{eqnarray*}\label{dderi}
\ell'(d)&=&-\frac{1}{m}(\sqrt{Ke^{-\lambda d}}+\frac{\sqrt{\log (4\gamma)}}{2})^2+1+\frac{\log (4\gamma)}{4m}\cr
\ell''(d)&=&\frac{\lambda}{m}(\sqrt{Ke^{-\lambda d}}+\frac{\sqrt{\log (4\gamma)}}{2})\sqrt{Ke^{-\lambda d}}>0.
\end{eqnarray*}
Thus $\ell(d)$ is a convex function on $[0,+\infty),$ and the optimal $d_{opt}$ value can be found by setting $\ell'(d)=0:$
\begin{eqnarray*}\label{eqn:d}
\sqrt{Ke^{-\lambda d}}=\sqrt{m+\frac{\log (4\gamma)}{4}}-\frac{1}{2}\sqrt{\log (4\gamma)}.
\end{eqnarray*}
This gives an unique optimal value
\begin{eqnarray}\label{dop1}
d_{opt} &=&\frac{1}{\lambda}\log \frac{K}{(\sqrt{m+\frac{1}{4}\log (4\gamma)}-\frac{1}{2}\sqrt{\log (4 \gamma)})^2}\\ \nonumber
&=&\frac{1}{\lambda}\left\{ \log  \frac{\left[\sqrt{m+\frac{1}{4}\log (4\gamma)}+\frac{1}{2}\sqrt{\log (4 \gamma)}\right]^2}{m} + \log \frac{K}{m} \right\},
\end{eqnarray}
which is equivalent to those in (\ref{d1}) under the assumption that $m = m(K) << \min(\log\gamma, K).$  Plugging $d= d_{opt}$ in (\ref{eqn:d}) back to (\ref{b1}) yields the choice of $b_{\gamma}$.

\subsection*{D. Proof of Theorem   \ref{thm6} }
By Theorems \ref{thm1} and \ref{thm1b}, the false alarm breakdown point  of our proposed method $N_{\alpha}$ can be found by finding the smallest $\epsilon$ value such that $I_{0}(\epsilon,\alpha;g_0) > 0$ for some distribution $g_0,$ where $I_{0}(\epsilon,\alpha; g_0)$ is defined in (\ref{conditionful00}). That is equivalent to 
\begin{eqnarray}\label{brkd}
\epsilon^*(N_{\alpha})= \inf \{\epsilon\ge 0: \underset{g_0}{\sup} \,  I_{0}(\epsilon,\alpha;g_0) > 0 \},
\end{eqnarray}

The remaining proof is based on a careful analysis of $I_{0}(\epsilon,\alpha;g_0)$  for any arbitrary outlier density function $g_0$.    For any $h_0(x)=(1-\epsilon)f_0(x)+\epsilon g_0(x)\in \hbar_{0,\epsilon},$ by  (\ref{conditionful00}),  we have
\begin{eqnarray}\label{equ:h0}
I_{0}(\epsilon,\alpha;g_0)  &=& - \frac{1-\epsilon}{1+\alpha}d_{\alpha}(f_{0},f_{1})+\epsilon \int (\frac{[f_{1}(x)]^{\alpha}-[f_{0}(x)]^{\alpha}}{\alpha})g(x) dx,
\end{eqnarray}
where $d_{\alpha}(f_{0},f_{1})$ is defined in  (\ref{eqndensitypower}) and is the density power divergence between $f_{0}$ and $f_{1}$ proposed by \cite{basu:1998}. Here we use the fact that $\int [f_{1}(x)]^{1+\alpha} d x = \int [f_{0}(x)]^{1+\alpha} d x$ when $f_0(x)$ and $f_1(x)$ come from the same location family.

By the definition of $M(\alpha)$ in (\ref{eqnM0M1}), it is clear from (\ref{equ:h0}) that
\begin{eqnarray}\label{muh}
\underset{g_0}{\sup}\, I_{0}(\epsilon,\alpha;g_0)
=-\frac{1-\epsilon}{1+\alpha}d_{\alpha}(f_{0},f_{1})+\epsilon M(\alpha).
\end{eqnarray}
Therefore, by (\ref{brkd}), if both $d_{\alpha}(f_{0},f_{1})$ and $M(\alpha)$ are finite,
the false alarm breakdown point of $N_{\alpha}$ should be
\begin{eqnarray}\label{conclu}
\epsilon^*(N_{\alpha}) = \frac{d_{\alpha}(f_{0},f_{1})}{d_{\alpha}(f_{0},f_{1})+(1+\alpha)M(\alpha)}.
\end{eqnarray}
If $d_{\alpha}(f_{0},f_{1})$ is finite but $M(\alpha)=+\infty$, by (\ref{brkd}) and (\ref{muh}), $\epsilon^*(N_{\alpha}) =0$.
If $d_{\alpha}(f_{0},f_{1})=+\infty$ but $M(\alpha)$ is finite, $\epsilon^*(N_{\alpha}) =1$. If both $d_{\alpha}(f_{0},f_{1})$ and $M(\alpha)$ are $+\infty$ and $\frac{d_{\alpha}(f_{0},f_{1})}{M(\alpha)}=\rho$, by (\ref{brkd}) and (\ref{muh}), we have $\epsilon^*(N_{\alpha})=\frac{\rho}{\rho+(1+\alpha)}$ no matter $\rho$ is finite or not.
Therefore, for all cases, the false alarm breakdown point of $N_{\alpha}$ have the same expression in (\ref{conclu}), which completes the proof of Theorem \ref{thm6}.



\bibliographystyle{IEEEtran}

\bibliography{changepoint}

\end{document}